\documentclass[reprint, twocolumn, showkeys, showpacs, preprintnumbers, amsmath, rsi, aip, longbibliography]{revtex4-1}

\usepackage{graphicx, color, bm, hyperref}

\hypersetup{
colorlinks=true,
urlcolor= blue,
citecolor=blue,
linkcolor= blue,
bookmarks=true,
bookmarksopen=false,
}

\pdfoutput=1

\begin{document}

\title{Etching of Cr tips for scanning tunneling microscopy of cleavable oxides}
\author{Dennis Huang}
\thanks{These authors contributed equally to this work.}
\author{Stephen Liu}
\thanks{These authors contributed equally to this work.}
\author{Ilija Zeljkovic}
\altaffiliation{Present address: Department of Physics, Boston College, Chestnut Hill, MA 02467, U.S.A.}
\affiliation{Department of Physics, Harvard University, Cambridge, MA 02138, U.S.A.}
\author{J. F. Mitchell}
\affiliation{Materials Science Division, Argonne National Laboratory, Argonne, IL 60439, U.S.A.}
\author{Jennifer E. Hoffman}
\email{jhoffman@physics.harvard.edu}   
\affiliation{Department of Physics, Harvard University, Cambridge, MA 02138, U.S.A.}
\date{\today}

\begin{abstract}
We report a detailed three-step roadmap for the fabrication and characterization of bulk Cr tips for spin-polarized scanning tunneling microscopy. Our strategy uniquely circumvents the need for ultra-high vacuum preparation of clean surfaces or films. First, we demonstrate the role of \textit{ex-situ} electrochemical etch parameters on Cr tip apex geometry, using scanning electron micrographs of over 70 etched tips. Second, we describe the suitability of the \textit{in-situ} cleaved surface of the layered antiferromagnet La$_{1.4}$Sr$_{1.6}$Mn$_2$O$_7$ to evaluate the spin characteristics of the Cr tip, replacing the UHV-prepared test samples that have been used in prior studies. Third, we outline a statistical algorithm that can effectively delineate closely-spaced or irregular cleaved step edges, to maximize the accuracy of step height and spin-polarization measurement.
\end{abstract}

\maketitle

\section{Introduction}

Spin-polarized scanning tunneling microscopy (SP-STM) is a powerful technique for real-space imaging of atomic-scale spin features.~\cite{Wiesendanger_RMP_2009, Bode_RPP_2003} Its implementation, starting from a conventional scanning tunneling microscope (STM) setup, requires careful preparation of (1) a tip with a well-defined magnetic termination and (2) a test sample with nanoscale magnetic structure.

Tips for SP-STM have been fabricated using bulk ferromagnetic (FM)~\cite{Cavallini_RSI_2000, Ceballos_SS_2003} or antiferromagnetic (AF)~\cite{Shvets_JAP_1992, LiBassi_APL_2007} materials, or by evaporating a thin magnetic film on a nonmagnetic tip.~\cite{Kubetzka_PRL_2002, Rodary_APL_2011} While FM tips afford larger spin contrast, AF tips produce negligible stray fields and are better suited for nondestructive imaging. AF tips etched from bulk Cr are one emerging candidate for SP-STM, ideal for their monatomic composition and high N\'{e}el temperature of 311 K.  They typically exhibit a canted tip magnetization which is rotatable in a 2 T field, sensitive to all 3D spatial components, and capable of atomic-resolution imaging.~\cite{LiBassi_APL_2007, Schlenhoff_APL_2010, Corbetta_JJAP_2012, Romming_Science_2013, Doi_PRB_2015} However, the extent to which the electrochemical preparation influences these characteristics is poorly understood. Systematic studies of tip etching parameters have been mostly limited to nonmagnetic W, whose electrochemistry differs notably from that of Cr.~\cite{Ibe_JVSTA_1990, Oliva_RSI_1996, Bastiman_JVSTB_2010}

A major practical advantage of bulk Cr tips is that they circumvent the need for complex ultra-high vacuum (UHV) cleaning and evaporation procedures, as well as \textit{in-situ} tip exchange.
However, the magnetic samples so far used to quantify spin polarization, e.g.\ Fe$_3$O$_4$,~\cite{Shvets_JAP_1992} Cr(001),~\cite{LiBassi_APL_2007} Fe/W(110),~\cite{Schlenhoff_APL_2010} and Co/Cu(111)~\cite{Corbetta_JJAP_2012}, all involve extensive surface preparation in a UHV environment.


Here we report a detailed roadmap for non-UHV Cr tip preparation and evaluation, using three key new strategies. First, we examine how \textit{ex-situ} fabrication parameters affect Cr tip apex geometry, which in turn influences both atomic- and spin-resolution imaging. We etched more than 70 Cr rods under various voltage sequences, and used a scanning electron microscope (SEM) to image the tips formed on both ends of the break junction. Second, we prepare test samples for tip characterization by mechanical cleaving as opposed to UHV cleaning and evaporation. We use the cuprate Bi$_2$Sr$_2$CaCu$_2$O$_{8+\delta}$ to evaluate atomic resolution, and the layered antiferromagnet La$_{1.4}$Sr$_{1.6}$Mn$_2$O$_7$ to evaluate spin polarization. Third, we introduce a Gaussian mixture model that can accurately quantify step heights and spin-polarization despite the common challenges of closely-spaced or irregular terraces. Our work charts a path to calibrated spin-polarized tunneling measurements, eliminating the need for UHV surface preparation tools.

\section{Chromium tips}

\begin{figure}
\includegraphics[scale=1]{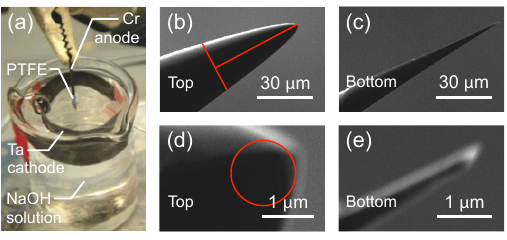}
\caption{(a) Photograph of etching setup. (b-e) Scanning electron micrographs of two Cr tips derived from the (b) top and (c) bottom rod ends of a single etch. Higher magnification images of the same tips are shown in (d, e). The perpendicular lines in (b) illustrate the procedure used in computing the aspect ratio (AR), and the circle in (d) illustrates the radius of curvature (RC).}
\label{Fig1}
\end{figure}

We etched tips from square 0.5 mm $\times$ 0.5 mm polycrystalline Cr rods (99.99$\%$ purity)~\footnote{The square 0.5 mm $\times$ 0.5 mm rods were cut from polycrystalline Cr foil purchased from Brooks Precision.} using the standard direct current (DC) drop-off method.~\cite{Chen_2007} Figure~\ref{Fig1}(a) gives a photograph of our setup. One end of the rod was covered with a 7 mm polytetrafluoroethylene (PTFE) tubing~\cite{Iijima_JJAP_1988} and immersed in a 5 M NaOH solution, such that the rod area in contact with the solution was minimized. Next, we applied a DC voltage to drive the anodic dissolution of Cr, eventually into CrO$_4$$^-$. The exposed portion of the rod was thinned until the weight below exceeded the tensile force and broke off, leaving behind a work-hardened tip on both ends of the break. The voltage was quickly shut off and both the remnant tip above the break (called top) and the remnant tip below the break (called bottom) were rinsed in deionized water and retained for subsequent examination.  We used a fresh solution (poured from the same stock) for each etch in order to standardize our tip preparation.

We evaluated the tips using a Zeiss Ultra Plus SEM. Figures~\ref{Fig1}(b-e) present sample micrographs of two tips derived from the bottom and top rod ends of a single etch, at two different magnifications. We utilized two metrics to assess the tip apex geometry: (1) Aspect ratio (AR), defined as length over width measured 50 $\mu$m from the tip end (perpendicular lines in Fig.~\ref{Fig1}(b)), and (2) radius of curvature (RC), computed from a polynomial fit to the tip apex contour (circle in Fig.~\ref{Fig1}(d)). Our results were robust across different definitions of AR with varying lengths from the tip end.

\begin{figure}
\includegraphics[scale=1]{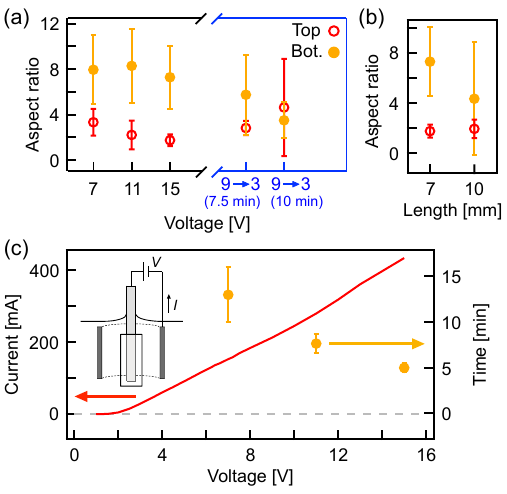}
\caption{(a) Plot of aspect ratio (AR) for several etch voltages, binned by top and bottom tips. Each point represents the average of 6 tips.  The right side of the axis break presents ARs associated with two-step etches -- first at 9 V for the indicated time duration (7.5 min or 10 min), then at 3 V until the drop-off. (b) Plot of average AR for two different lengths of PTFE tubing, corresponding to a 43$\%$ difference in the combined weight of the Cr rod and tubing.  Each point represents the average of 7 tips, etched at 15 V. (c) Potentiostatic polarization ($I$--$V$) curve (solid line) for our given etching setup (inset schematic).  The onset of current near 2 V corresponds to the negative cell potential, which is dependent on the setup geometry.  The circles denote corresponding average etch times.}
\label{Fig2}
\end{figure}

Figure~\ref{Fig2}(a) displays the average tip apex AR for etch voltages of 7 V, 11 V and 15 V, binned by top and bottom tips.  The bottom Cr tips are statistically sharper than their top counterparts, because they were instantaneously disconnected at the break and did not sustain residual etching in the few seconds before the voltage was manually shut off.~\cite{Bryant_RSI_1987, Iijima_JJAP_1988, Ceballos_SS_2003} Furthermore, the average ARs are largely uncorrelated with the etch voltage or sequence.  No improvements are detected with a two-step process, whereby the voltage was reduced from 9 V to 3 V after a fixed time period (right end of axis break in Fig.~\ref{Fig2}(a)).~\cite{LiBassi_APL_2007} In fact, the two-step process, with its longer etch time typically exceeding 30 minutes, was likely more susceptible to external perturbations (e.g. vibrations or solution evaporation), resulting in larger AR variability and unclear distinction between top and bottom tips. In the one-step processes, we also find the average RC of the bottom tips to be smaller than that of the top tips, but the trend is smaller and lacks the statistical significance seen in the AR.

We also considered the effect of the rod weight below the break junction on the tip apex geometry.~\cite{Ibe_JVSTA_1990} Figure~\ref{Fig2}(b) presents a comparison between a set of seven Cr rods covered with 7 mm long PTFE tubing and another set of seven covered with 10 mm long tubing, representing a 43$\%$ increase in combined weight (rod and tubing).  The increased weight yields greater variability in the bottom tip AR despite an average 23$\%$ faster etch time, possibly because the larger weight induced an earlier and more uncontrolled break.

To further understand the role of the etch voltage, we obtained a potentiostatic polarization ($I$--$V$) curve for our given setup, shown in Fig.~\ref{Fig2}(c) (solid line). A minimum of $\sim$2 V corresponding to the negative cell potential is required to drive a measurable reaction current. Above this threshold, the current rises approximately linearly with voltage, up to 15 V. Unlike previous reports on W tips,~\cite{Ibe_JVSTA_1990, Oliva_RSI_1996} we do not observe any saturation or upturn in the current that may be indicative of competing or secondary reactions. Figure~\ref{Fig2}(c) also depicts average etch times (circles) for the voltage settings used in Fig.~\ref{Fig2}(a), which are faster with increasing current.  Taken together, Fig.~\ref{Fig2} suggests that for Cr tips, larger voltages may be used to decrease etch times without affecting tip sharpness or inducing additional reactions.

\section{Scanning Tunneling Microscopy}

\begin{figure}
\includegraphics[scale=1]{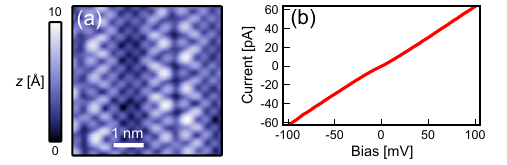}
\caption{Topographic image of Bi$_2$Sr$_2$CaCu$_2$O$_{8+\delta}$ taken with a Cr tip. Setpoint: 100 mV, 20 pA; $T$ = 7 K.
(b) Linear $I$-$V$ spectrum of polycrystalline Au taken with a Cr tip, suggesting a featureless tip density of states. This is the same Cr tip used to acquire the La$_{1.4}$Sr$_{1.6}$Mn$_2$O$_7$ data in Fig.~\ref{Fig5}. Setpoint: -100 mV, 1.5 G$\Omega$ junction resistance; $T$ = 6.6 K.}
\label{Fig3}
\end{figure}

Prior to use for STM imaging, the Cr tips were cleaned by field emission onto a Cr or Au foil within the STM. Typically, we bring the tip into constant-current feedback with a setpoint of 50-100 V and 0.5-3 $\mu$A for several minutes. This removes any oxides and restructures the terminal atoms on the tip apex, which allows atomic-resolution tunneling largely independent of the post-etch RC. A high post-etch AR, however, is still necessary to ensure a small RC after repeated field emission attempts, and to probe surfaces with large corrugations and step edges. To demonstrate the spatial resolution capabilities of our Cr tips, we used the archetypical cuprate Bi$_2$Sr$_2$CaCu$_2$O$_{8+\delta}$. Figure~\ref{Fig3}(a) shows a 5 nm $\times$ 5 nm topographic image of Bi$_2$Sr$_2$CaCu$_2$O$_{8+\delta}$ taken with a Cr tip at 7 K. Both the unit cell and the structural supermodulation can be seen. Another purpose of the field emission is to ensure that our Cr tips exhibit flat tunneling conductance on simple metallic Au, as shown in Fig.~\ref{Fig3}(b). This suggests a featureless tip density of states, consistent with theory.~\cite{Passoni_PRB_2009}

\begin{figure}
\includegraphics[scale=1]{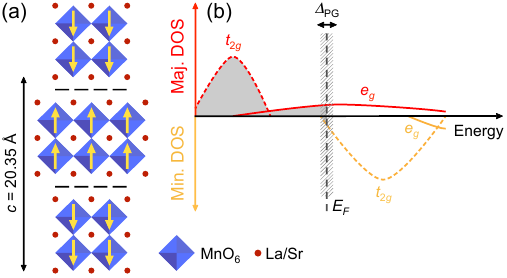}
\caption{(a) Crystal structure of La$_{2-2x}$Sr$_{1+2x}$Mn$_2$O$_7$, depicting the interbilayer antiferromagnetic (AF) order at $x$ = 0.30. Horizontal dashed lines mark the expected cleavage planes, spaced $c$/2 units apart. (b) Schematic diagram of the projected Mn $3d$ density of states (DOS).  The shading indicates the populated states below the Fermi energy.  For $0.30 \lesssim x \lesssim 0.40$, the system is close to a half-metallic ferromagnet (intrabilayer), with possibly an electron pocket of minority spin $t_{2g}$ character.  The hatching represents a ``pseudogap" detected by angle-resolved photoemission spectroscopy and scanning tunneling microscopy.~\cite{Ronnow_Nat_2006, Loviat_Nanotech_2007, Massee_NatPhys_2011}}
\label{Fig4}
\end{figure}

To circumvent the need for UHV preparation of magnetic test sample to evaluate the spin sensitivity of the Cr tip, we propose the use of the cleavable bilayer manganite La$_{2-2x}$Sr$_{1+2x}$Mn$_2$O$_7$. In this material, strong coupling of spin, orbital and lattice degrees of freedom underlies a colossal magnetoresistance effect,~\cite{Kimura_Science_1996} as well as a diverse display of magnetic orders.~\cite{Li_PRB_1999, Mitchell_JAP_1999, Mitchell_JPCB_2001} Figure~\ref{Fig4}(a) shows a schematic structure of La$_{2-2x}$Sr$_{1+2x}$Mn$_2$O$_7$ in the Ruddlesden-Popper phase. Sr substitution alters the Mn$^{3+}$/Mn$^{4+}$ valency and rapidly changes the magnetic ground state by a delicate tuning of double exchange and crystal field effects.~\cite{Welp_JAP_2001} At $x$ = 0.30 (and below 90 K), spins within a given bilayer are aligned parallel to the $c$-axis, but antiparallel to spins in the adjacent bilayers (Fig.~\ref{Fig4}(a)).~\cite{Li_PRB_1999, Mitchell_JAP_1999, Welp_JAP_2001, Mitchell_JPCB_2001} If cleavage in the $a$-$b$ plane generates step edges spanning adjacent bilayers, then the AF coupling between terraces can be observed using SEM.~\cite{Konoto_PRL_2004} Furthermore, we expect the spin contrast signal to be large due to the approximate half-metallic ferromagnetism within each bilayer. Figure~\ref{Fig4}(b) shows a schematic diagram of the projected Mn 3$d$ density of states that dominate the La$_{2-2x}$Sr$_{1+2x}$Mn$_2$O$_7$ electronic structure near the Fermi energy. For $0.30 \lesssim x \lesssim 0.40$, the occupied bands within $\sim 2$ eV of the Fermi energy carry majority spin,~\cite{Huang_PRB_2000} except for a small electron pocket of minority spin $t_{2g}$ character that may be present at the Brillouin zone center.~\cite{Saniz_PRL_2008, Sun_SRep_2013}

\begin{figure}
\includegraphics[scale=1]{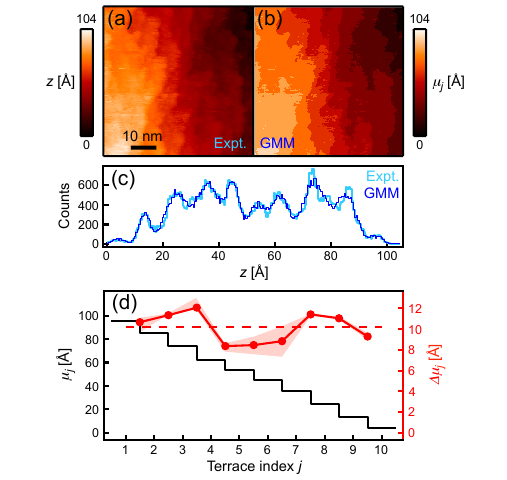}
\caption{(a) Topographic image of La$_{1.4}$Sr$_{1.6}$Mn$_2$O$_7$, taken with a Cr tip. Setpoint: $-$250 mV, 15 pA; $T$ = 6.5 K. (b) Mean values ($\mu_j$) of the terrace heights, identified through statistical inference with a Gaussian mixture model (GMM). (c) Validation of the GMM through a comparison of its simulated histogram (dark blue) with the actual histogram of (a) (light blue). (d) Plots of $\mu_j$ values (black) and their differences, $\Delta \mu_j = \mu_j - \mu_{j+1}$ (red). The dashed red line marks the average height difference of $c$/2 = 10.175 \AA. The red shading depicts uncertainties in $\mu_j$ due to background tilt corrections.}
\label{Fig5}
\end{figure}

In Fig.~\ref{Fig5}(a), we present a constant-current image of cold-cleaved La$_{1.4}$Sr$_{1.6}$Mn$_2$O$_7$, obtained with a Cr tip at 6.5 K.~\footnote{The Cr tips used in Figs.~\ref{Fig3},~\ref{Fig5} were the top halves of rods etched at 10 V for $\sim$7 minutes, then at 3 V until the drop-off. These tips were rinsed in deionized water post-etching, but were not SEM-imaged, to avoid possible contamination.} We make two remarks. First, due to inferred in-plane screening, STM does not resolve atomic-scale features on the surface of La$_{2-2x}$Sr$_{1+2x}$Mn$_2$O$_7$, save for occasional nanometer-sized patches of square lattice corrugations ascribed to trapped polarons.~\cite{Ronnow_Nat_2006, Santis_JSNM_2007} It has also been suggested that mobile oxygen defects obscure atomic-resolution tunneling in the layered manganites.~\cite{Bryant_NatComm_2011} Second, the cleavage is expected to occur between La/SrO buffer planes (Fig.~\ref{Fig4}(a)), as deduced from X-ray photoelectron spectroscopy,~\cite{Loviat_Nanotech_2007} and complemented by other non-spin-polarized STM studies that consistently found step edge heights to be integer multiples of the half-unit cell ($c$/2).~\cite{Ronnow_Nat_2006, Loviat_Nanotech_2007, Massee_NatPhys_2011} Here, multiple terraces are evident over a 60 nm $\times$ 60 nm area, and we instead observe small variations in their height differences.

To assess whether these apparent height variations could originate from spin-polarized tunneling, we performed statistical inference with a Gaussian mixture model (GMM), commonly employed in clustering applications.~\cite{Murphy_ML_2012} The model assumes there are $k$ (= 10) underlying terraces, indexed by $j$, each occupying a fraction $\phi_j$ of the field of view. Within each terrace, the heights are Gaussian-distributed with mean $\mu_j$ and standard deviation $\sigma_j$. The objective is to find values of $\phi_j$, $\mu_j$, and $\sigma_j$ that maximize the (logarithm of the) probability for which the GMM can instantiate the actual data (see supplementary material). This problem can be solved iteratively through the expectation-maximization algorithm.~\cite{Dempster_JRSS_1977} We note the general utility of the GMM approach for STM identification of irregular terrace boundaries by optimization, rather than manual assignment.

Figure~\ref{Fig5}(b) reveals the underlying terraces inferred from the GMM. Our procedure is validated by the close match of histograms generated from the actual data and the GMM (Fig.~\ref{Fig5}(c)), which share a Pearson correlation coefficient of $r$ = 0.98. The mean terrace heights $\mu_j$ and their differences $\Delta \mu_j = \mu_j - \mu_{j+1}$ are plotted in Fig.~\ref{Fig5}(d). In the case of bilayer terraces with AF coupling, we expect $\Delta \mu_j$ to exhibit bimodal switching about $c$/2,~\cite{Wiesendanger_PRL_1990} due to a spin-valve contribution to the tunneling current that depends on the cosine of the angle subtended by the tip and sample magnetizations. Here, the height differences may follow a bimodal distribution, but they do not alternate as expected. We enumerate a few possible reasons: (1) A complication could arise from occasional insertions of ferromagnetically-coupled terraces. Although adjacent bilayers have antiparallel spins in the $x$ = 0.30 compound, their spins become aligned as soon as the local dopant concentration is raised to $x$ = 0.32.~\cite{Mitchell_JAP_1999, Mitchell_JPCB_2001} In our case, the nanometer widths of our terraces could be comparable to the length scale of dopant inhomogeneity. We note that the $c$-axis change within $0.30 \le x \le 0.40$, which is around 2\%, is too small to explain the observed variations in $\Delta \mu_j$.~\cite{Okuda_MSEB_1999} (2) Previous works have shown that La$_{2-2x}$Sr$_{1+2x}$Mn$_2$O$_7$ crystals in the range $0.36 \le x \le 0.50$ can develop a 1-nm-thick layer of nonmagnetic insulator at their surfaces.~\cite{Freeland_NatMat_2004}. However, spin-polarized SEM measurements with penetration depth $\le$ 1 nm have detected layered AF texture in the $x = 0.30$ compound.~\cite{Konoto_PRL_2004} (3) Stacking faults can occur during the growth of bilayer crystals, but previous STM studies showed that they were rare and possessed a distinct topographic signal.~\cite{Massee_NatPhys_2011} Overall, our use of the GMM to extract spin signals is easy to implement and widely applicable, but further measurements, preferably with wider terraces, are needed to fully characterize the magnetic textures of our La$_{1.4}$Sr$_{1.6}$Mn$_2$O$_7$ samples and Cr tips.

\section{Summary}

In summary, we detailed simple approaches to quantitative, atomic-resolution SP-STM that do not require UHV preparation conditions. First, we investigated the preparation of bulk Cr tips by DC drop-off etching. Our findings indicate that the bottom tips are statistically sharper than their top counterparts, and large voltages for faster etches do not reduce tip AR or produce additional reactions. Second, we tested the spatial and spin resolutions of our Cr tips on \textit{in-situ}-cleaved crystals. We demonstrated atomic-resolution imaging on Bi$_2$Sr$_2$CaCu$_2$O$_{8+\delta}$ and flat conductance on Au. More investigation is needed to fully quantify the layered magnetic texture of La$_{1.4}$Sr$_{1.6}$Mn$_2$O$_7$; nevertheless, our use of the GMM demonstrates a rigorous framework wherein mean terrace heights can be extracted in the presence of complications such as in-plane screening or possible mobile defects. Our work may aid applications of quantitative SP-STM to cleaved planes of quantum materials that are grown as high-quality single crystals, such as cuprate and Fe-based superconductors, colossal magnetoresistance materials, and topological insulators. Local spin mapping is needed to unravel the exotic electronic behavior of these materials.

\section{Supplementary Material}

See supplementary material for additional images of the etching setup and procedure, a review of the electronic structure of La$_{2-2x}$Sr$_{1+2x}$Mn$_2$O$_7$, details of the Gaussian mixture model used to delineate step edges, and details of the spin-polarization calculation.

\section{Acknowledgements}

We thank Genda Gu for providing the Bi$_2$Sr$_2$CaCu$_2$O$_{8+\delta}$ sample imaged in this work. Work at Harvard was supported by the National Science Foundation under Grant.\ No.\ DMR-0847433. D.H. acknowledges support from an NSERC PGS-D fellowship.
Work at Argonne National Laboratory (crystal growth and characterization) is supported by the U.S. Department of Energy, Office of Science, Basic Energy Sciences, Materials Science and Engineering Division.

\onecolumngrid
\newpage

\section*{Supplementary Material}

\setcounter{figure}{0}
\setcounter{equation}{0}
\setcounter{table}{0}
\setcounter{section}{0}
\makeatletter
\renewcommand{\thefigure}{S\@arabic\c@figure}
\renewcommand{\theequation}{S\@arabic\c@equation}
\renewcommand\theHfigure{S\arabic{figure}}
\renewcommand{\thetable}{S\@arabic\c@table}

\section{Etching Setup and Procedure}

\begin{figure}[h]
\includegraphics[scale=1]{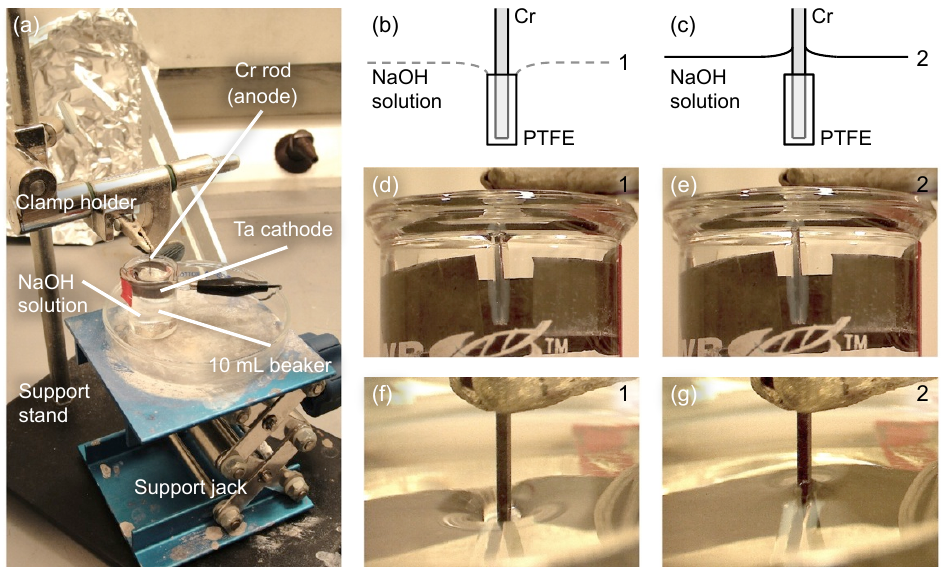}
\caption{(a) Larger scale photograph of the etching setup. (b, c) Schematic and (d-g) photographic demonstration of a two-step rod immersion procedure. The figures labeled ``1" (b, d, f) depict the initial entry of the bare Cr rod into the NaOH solution, during which the meniscus is bent towards the top of the polytetrafluoroethylene (PTFE) tubing. The figures labeled ``2" (c, e, g) depict the second step, wherein the solution level is gradually jacked up until the instant the meniscus surpasses the PTFE tubing and re-forms around the rod. This step is carried out in order to confine the etching activity to a small region above the PTFE tubing.}
\label{FigS1}
\end{figure}

Figure~\ref{FigS1} provides additional photographic documentation of the etching setup and procedure used in this work. In Fig.~\ref{FigS1}(a), we present a larger scale image of our setup.  A support jack raises a 10 mL beaker with 5 M NaOH solution towards a Cr rod, fixed in place by a clamp holder attached to a support stand.  We use a set square [pictured in Fig.~\ref{FigS1}(a)] to ensure that the Cr rod axis is orthogonal to the solution interface.  We also carefully cut the polytetrafluoroethylene (PTFE) tubing with a razor blade so that its top end is flat and minimally deformed from a circular cross-section.  Its bottom end is pinched to prevent it from slipping off the Cr rod.  To drive the etching reaction, we use a DC-regulated power supply (Tenma 72-6628).  A 91 $\Omega$ resistor placed in series with the circuit stabilizes the etch speed.

We use a two-step rod immersion procedure to minimize the macroscopic tip aspect ratio, in order to reduce vibrations when scanning. Figures~\ref{FigS1}(b-g) provide multiple depictions of each step.  

\section{Electronic Structure of L\lowercase{a}$_{2-2x}$S\lowercase{r}$_{1+2x}$M\lowercase{n}$_2$O$_7$}

\begin{figure}[h]
\includegraphics[scale=1]{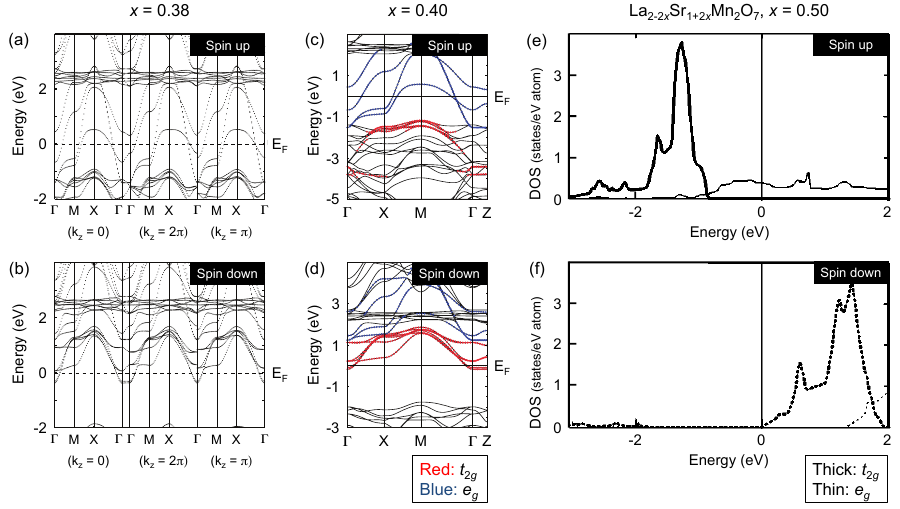}
\caption{(a, b) Majority and minority spin bands calculated for La$_{2-2x}$Sr$_{1+2x}$Mn$_2$O$_7$, $x = 0.38$, using the all-electron full-potential Korringa-Kohn-Rostoker (KKR) and linearized augmented plane-wave (LAPW) methods.  Reprinted from~\citet{Sun_SRep_2013} (c, d) Majority and minority spin bands calculated for the $x = 0.40$ compound, using the all-electron full-potential LAPW method with the generalized gradient approximation (GGA).  The red (blue) circles represent states of dominant $t_{2g}$ ($e_g$) character. Reprinted from~\citet{Saniz_PRL_2008}. (e, f) Spin-resolved projected Mn 3$d$ density of states (DOS) for the $x = 0.50$ compound, calculated using the full-potential linear muffin-tin orbital (FLMTO) method.  The thick (thin) lines represent the $t_{2g}$ ($e_g$) states.  Reprinted from~\citet{Huang_PRB_2000}}
\label{FigS2}
\end{figure}

The schematic electronic band structure of La$_{2-2x}$Sr$_{1+2x}$Mn$_2$O$_7$ presented in Fig. 4(b) of the main text is informed by calculations and experiments in literature.  Figures~\ref{FigS2}(e, f) display the projected Mn 3$d$ density of states (DOS) for the $x=0.50$ compound, calculated by \citet{Huang_PRB_2000}  Their results indicate that  LaSr$_{2}$Mn$_2$O$_7$ is a half-metallic ferromagnet with a band gap in the minority spin channel of $1.7-1.9$ eV.  Increasing electron doping (decreasing $x$) is modeled in the rigid band approximation, resulting in an upward shift of the Fermi level to populate some minority spin $t_{2g}$ states.  The rigid band shift is confirmed by angle-resolved photoemission spectroscopy (ARPES) and $ab$-$initio$ calculations for the $x = 0.38$ (Figs.~\ref{FigS2}(a, b), reproduced from \citet{Sun_SRep_2013}) and $x = 0.40$ (Figs.~\ref{FigS2}(c, d), reproduced from \citet{Saniz_PRL_2008}) compounds, which predict an electron pocket of minority spin $t_{2g}$ character around the $\Gamma$ point.

An additional feature in the electronic structure of La$_{2-2x}$Sr$_{1+2x}$Mn$_2$O$_7$ is a ``pseudogap" detected by ARPES~\cite{Dessau_PRL_1998, Mannella_Nat_2005} and scanning tunneling microscopy (STM).~\cite{Ronnow_Nat_2006, Loviat_Nanotech_2007, Massee_NatPhys_2011}  For the $x = 0.30$ compound, STM extracts a gap magnitude of $196 \pm 12$ meV by fitting the temperature-dependent zero-bias conductivity to a thermal activation model.~\cite{Ronnow_Nat_2006} Such a pseudogap is not the subject of our discussion, but we indicate its presence in Fig. 4(b) of the main text for completeness.

\section{Gaussian Mixture Model}

We detail our statistical procedure for extracting mean terrace heights from our topographic image of La$_{1.4}$Sr$_{1.6}$Mn$_2$O$_7$.

\textit{Step 1. Gaussian mixture model (GMM)} - For our given topography (Fig. 5(a) of main text), we assume there are $k$ (= 10) underlying terraces, indexed by $j$, each occupying a fraction $\phi_j$ of the field of view. Within each terrace, the heights are Gaussian-distributed with mean $\mu_j$ and standard deviation $\sigma_j$. 

\textit{Step 2. Inference} - We determine $\mu_j$, $\sigma_j$, and $\phi_j$ by maximizing the logarithm of the probability for which the GMM can instantiate our given topography $z(x, y)$;~\cite{Murphy_ML_2012} i.e., we find
\begin{align}
& \underset{\phi_j, \mu_j, \sigma_j}{\textrm{argmax}} \log p(\textrm{topo.} | \textrm{GMM with params. }\phi_j, \mu_j, \sigma_j) \nonumber \\
\label{Eqloglikelihood}
& = \underset{\phi_j, \mu_j, \sigma_j}{\textrm{argmax}} \log \prod_{x, y} \bigg[ \sum_{j=1}^k  \frac{\phi_j}{\sqrt{2 \pi} \sigma_j} \exp \bigg[ -\frac{(z(x,y)-\mu_j)^2}{2 \sigma_j^2}\bigg] \bigg].
\end{align}
Since Eq.~\ref{Eqloglikelihood} has no closed-form solution, we use the iterative expectation-maximization algorithm~\cite{Dempster_JRSS_1977} to find $\mu_j$, $\sigma_j$, and $\phi_j$. Optimal values are given in Table~\ref{Tab1}.

\begin{table}[h]
\center
\setlength{\tabcolsep}{10pt}
\begin{tabular}{c|ccc}
\hline
$j$ & $\mu_j$ [\AA] & $\sigma_j$ [\AA] & $\phi_j$ \\
\hline\hline
1 & 95.75 $\pm$ 0.09 & 2.02 $\pm$ 0.06 & 0.01 \\
2 & 85.08 $\pm$ 0.05 & 3.47 $\pm$ 0.04 & 0.12 \\
3 & 73.73 $\pm$ 0.05 & 4.21 $\pm$ 0.04 & 0.19 \\
4 & 61.64 $\pm$ 0.05 & 3.00 $\pm$ 0.05 & 0.10 \\
5 & 53.28 $\pm$ 0.07 & 3.42 $\pm$ 0.06 & 0.10 \\
6 & 44.80 $\pm$ 0.04 & 2.28 $\pm$ 0.04 & 0.09 \\
7 & 35.96 $\pm$ 0.06 & 4.25 $\pm$ 0.05 & 0.19 \\
8 & 24.54 $\pm$ 0.05 & 3.76 $\pm$ 0.05 & 0.14 \\
9 & 13.49 $\pm$ 0.04 & 2.04 $\pm$ 0.04 & 0.04 \\
10 & 4.18 $\pm$ 0.10 & 2.02 $\pm$ 0.08 & 0.01\\
\hline
\end{tabular}
\caption{Optimized parameters for the Gaussian mixture model: $j$ is the terrace index, $\mu_j$ is the mean terrace height, $\sigma_j$ is the standard deviation of pixels within each terrace, and $\phi_j$ is the fraction of the field of view occupied by each terrace.}
\label{Tab1}
\end{table}

\begin{figure}[h]
\includegraphics[scale=1]{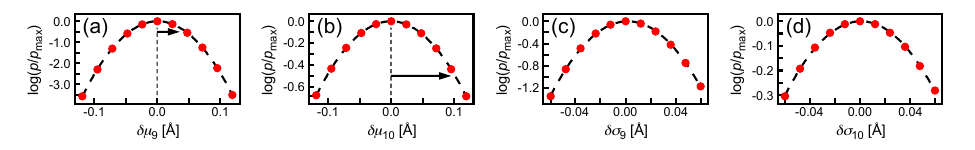}
\caption{Analysis of parameter uncertainties from the GMM. The logarithm probability ($\log p$) drops quadratically from its maximum value ($\log p_{\textrm{max}}$) as either $\mu_j$ or $\sigma_j$ are varied by increments $\delta \mu_j$ or $\delta \sigma_j$ away from their optimal values. We show four such examples: (a) $\mu_9$, (b) $\mu_{10}$, (c) $\sigma_9$, and (d) $\sigma_{10}$. From a quadratic fit to the computed logarithm probabilities, we can extract a Gaussian uncertainty, which corresponds to the position at which $\log (p/p_{\textrm{max}})$ falls to $-0.5$ [arrows in (a, b)].}
\label{FigS3}
\end{figure}

To determine the uncertainties associated with $\mu_j$ and $\sigma_j$, we separately vary each parameter while holding the others fixed at their optimal values. In every case, four of which are shown in Fig.~\ref{FigS3}, the logarithm probability drops quadratically, indicating that the uncertainties are Gaussian. To determine the uncertainties associated with $\phi_j$, we additionally require a Lagrange multiplier enforcing $\sum_{j=1}^k \phi_j = 1$. For simplicity, we have not performed this analysis.

\begin{figure}[h]
\includegraphics[scale=1]{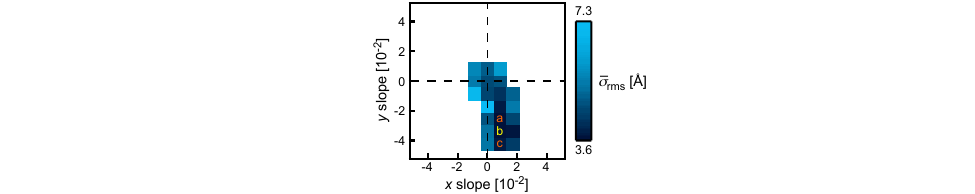}
\caption{Optimization of background plane subtraction for Fig. 5 of the main text. The pixel labeled ``b'' marks the $x$ and $y$ slopes that minimize the root-mean-square standard deviation $\bar{\sigma}_{\textrm{rms}}$ of all the terraces. Neighboring pixels labeled ``a'' and ``c'' are used to determine the uncertainties in $\mu_j$ due to tilt corrections.}
\label{FigS4}
\end{figure}

\textit{Step 3. Background plane subtraction} - To correct a small background tilt of our sample relative to the STM tip scanner, we subtract a plane from our topographic image. The $x$ and $y$ slopes are chosen so to minimize 
\begin{equation}
\bar{\sigma}_{\textrm{rms}} = \bigg[ \sum_{j=1}^k \phi_j \sigma_j^2 \bigg]^{1/2},
\end{equation}
the root-mean-square standard deviation across all terraces [Fig.~\ref{FigS4}].  

The uncertainty in resolving the background tilt produces much larger errors in $\mu_j$ compared to those derived from statistical inference with the GMM (Table~\ref{Tab1} and Fig.~\ref{FigS3}). To assess the errors from the former, we examine $\mu_j$ values for two sets of tilt corrections that deviate slightly from the optimal background plane (``a'' and ``c'' in Fig.~\ref{FigS4} and Table~\ref{Tab2}). These variations are represented by the faint red background of Fig. 5(d) of the main text.

\begin{table}[h]
\center
\setlength{\tabcolsep}{10pt}
\begin{tabular}{c|ccc}
\hline 
($x$ slope, $y$ slope) & a. $(0.008, -0.025)$ & \textbf{b.} $\textbf{(0.008, -0.034)}$ & c. $(0.008, -0.042)$ \\
$\bar{\sigma}_{\textrm{rms}}$ [\AA] & 3.623 & \textbf{3.619} & 3.802\\
\hline\hline
$\Delta \mu_1$ [\AA] & 11.03 & \textbf{10.68} & 9.69 \\
$\Delta \mu_2$ [\AA] & 11.52 & \textbf{11.35} & 11.41 \\
$\Delta \mu_3$ [\AA] & 13.01 & \textbf{12.08} & 11.93 \\
$\Delta \mu_4$ [\AA] & 8.58 & \textbf{8.36} & 7.83 \\
$\Delta \mu_5$ [\AA] & 9.32 & \textbf{8.48} & 7.53 \\
$\Delta \mu_6$ [\AA] & 10.23 & \textbf{8.84} & 7.34 \\
$\Delta \mu_7$ [\AA] & 11.48 & \textbf{11.42} & 11.26 \\
$\Delta \mu_8$ [\AA] & 11.23 & \textbf{11.05} & 10.97 \\
$\Delta \mu_9$ [\AA] & 9.66 & \textbf{9.31} & 8.96 \\
\hline
\end{tabular}
\caption{Mean terrace height differences ($\Delta \mu_j = \mu_j - \mu_{j+1}$) for three sets of tilt corrections (labeled ``a,'' ``b,'' and ``c'' in Fig.~\ref{FigS4}). The bolded values ``b'' correspond to the optimal tilt correction.}
\label{Tab2}
\end{table}

\bigskip \textit{Step 4. Terrace identification} - Each pixel $(x, y)$ has a probability $w_j(x, y)$ of belonging to terrace $j$, given by
\begin{equation}
w_j(x, y) = C \frac{\phi_j}{\sqrt{2 \pi} \sigma_j} \exp \bigg[ -\frac{(z(x,y)-\mu_j)^2}{2 \sigma_j^2}\bigg],
\end{equation}
where $C$ is a normalization constant. To produce Fig. 5(b) of the main text, we substitute all the pixels in Fig. 5(a) with the mean height of the terrace with which they are most probably associated.

\textit{Step 5. Simulated histogram} - To validate our GMM, we simulate a histogram in Fig. 5(c) of the main text, using the optimal values of $\mu_j$, $\sigma_j$, and $\phi_j$ in Table~\ref{Tab1}. We draw $N$ = 65536 samples, the total number of pixels in our topography. First, we randomly assign each sample to one of the $k$ terraces, with probability $\phi_j$ for terrace $j$. Second, for a pixel assigned to terrace $j$, we randomly determine its height according to a Gaussian probability distribution with mean $\mu_j$ and standard deviation $\sigma_j$. 

\section{Energy-integrated junction polarization}

In the case of spin-polarized tunneling between a Cr tip and antiferromagnetic (AF) terraces in La$_{1.4}$Sr$_{1.6}$Mn$_2$O$_7$, we expect $\Delta \mu_j$ to exhibit bimodal switching between $c/2 + \Delta z$ and $c/2 - \Delta z$, due to a spin-valve contribution to the tunneling current that depends on the cosine of the angle subtended by the tip and sample magnetizations. From these apparent height variations $\Delta z$, we can extract an energy-integrated junction polarization $P_{\perp}$.  

In the Bardeen formalism,~\cite{Bardeen_PRL_1961} the spin-polarized tunneling conductance (at $T$ = 0) is given by 
\begin{equation}\label{EqS1}
\frac{dI}{dV} = 2\pi^2 G_{0}|M_0|^2(\rho_s\rho_t+m_sm_t\cos\theta),
\end{equation}
where $\rho_{s,t} = \rho_{s,t}^{\uparrow}+\rho_{s,t}^{\downarrow}$ is the total density of states of the sample/tip, $m_{s,t} = \rho_{s,t}^{\uparrow}-\rho_{s,t}^{\downarrow}$ is the spin-polarized difference, $\theta$ is the angle subtended by the sample and tip magnetizations, $G_0 = \frac{2e^2}{h}$ is the conductance quantum, and $M_0 \propto e^{-\frac{\sqrt{2m\Phi}}{\hbar} z}$ is the matrix element ($\Phi$ is the local barrier height).~\cite{Chen_2007} Assuming a sample bias of $-V$, the spin-polarized tunneling current can be written as
\begin{equation}\label{EqS2}
I = I_0(1+P\cos\theta),
\end{equation}
where $I_0$ is a non-magnetic contribution to the current and  
\begin{equation}\label{EqS3}
P = \frac{\int_{-eV}^{0} \! m_{s}(\varepsilon) m_{t}(\varepsilon+eV) d\varepsilon}{\int_{-eV}^{0} \! \rho_{s}(\varepsilon) \rho_{t}(\varepsilon+eV) d\varepsilon}
\end{equation}
represents a convolution of the sample and tip magnetizations. In obtaining Eq.~\ref{EqS2}, we have assumed that $M_0$ and $\cos\theta$ are independent of energy over the range considered. 

Next, the tunneling current between a Cr tip and AF terraces in La$_{1.4}$Sr$_{1.6}$Mn$_2$O$_7$ is given by $I_{\uparrow\uparrow} = I_0(1+P\cos\theta)$ when the out-of-plane components of the sample and tip magnetizations are parallel, and $I_{\uparrow\downarrow} = I_0(1-P\cos\theta)$ when they are antiparallel (here, $\theta$ is the tip magnetization angle relative to the surface normal).  In constant-current feedback mode, this translates to a logarithmic increase $+\Delta z_1$ or decrease $- \Delta z_2$ in the tip-sample separation $z$, such that $I_{\uparrow\uparrow}(z+\Delta z_1)=I_{\uparrow\downarrow}(z-\Delta z_2)$.  Solving for $P\cos\theta$ yields 
\begin{equation}
P_{\perp} = P\cos\theta =\frac{e^{\frac{\sqrt{8m\Phi}}{\hbar}\Delta z}-1}{e^{\frac{\sqrt{8m\Phi}}{\hbar}\Delta z}+1},
\end{equation}
where $\Delta z = \Delta z_1 + \Delta z_2$ and $\Phi$ is approximated by the average of the tip and sample work functions.~\cite{Wiesendanger_PRL_1990}  

\begin{figure}[h]
\includegraphics[scale=1]{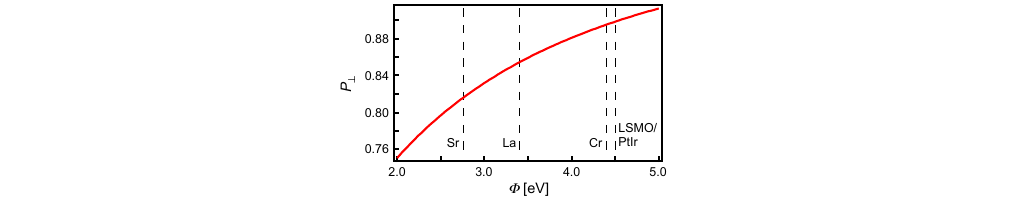}
\caption{Theoretical energy-integrated junction polarization $P_{\perp}$ as a function of the barrier height $\Phi$. A $\Delta z$ value of 1.35 \AA~was assumed. Dashed lines mark work functions of constituent elements.~\cite{Lange_2005} The ``LSMO/PtIr'' value was extracted from a non-spin-polarized tunneling measurement of a La$_{2-2x}$Sr$_{1+2x}$Mn$_2$O$_7$ compound with $x = 0.32$, taken with a PtIr tip at 20 K.~\cite{Huang_2007}}
\label{FigS5}
\end{figure}

As a demonstration of this procedure, Fig.~\ref{FigS5} shows the theoretical $P_{\perp}$ as a function of $\Phi$. Work functions of constituent elements are shown for reference. We take $\Delta z$ to be the standard deviation of the $\Delta \mu_j$ values in Fig. 5(d) of the main text (also listed in Table~\ref{Tab2}), but note that $\Delta \mu_j$ does not exhibit the clear bimodal switching expected from spin-polarized tunneling.   

\end{document}